\documentstyle[11pt,newpasp,twoside,epsf,psfig]{article}
\def\edcomment#1{\iffalse\marginpar{\raggedright\sl#1\/}\else\relax\fi}
\reversemarginpar

\begin{document}
\title{High Redshift Galaxy Clusters as Probes of Cosmology}
 \author{Jesper Sommer-Larsen and Martin G\"otz}
\affil{Theoretical Astrophysics Center, Juliane Maries Vej 30,\\
DK-2100 Copenhagen \O, Denmark}

\begin{abstract}
A few, simple and qualitative examples of the potential of galaxy
clusters as diagnostics of cosmology are presented. In relation to these we
discuss briefly three ongoing or forthcoming cluster surveys in the 
optical/NIR, X-rays and the cosmic microwave radiation background. 
\end{abstract}
\vspace{-5mm}
\section{Introduction}
There is continuing evidence that the first Doppler peak in the angular
power spectrum of the Cosmic Microwave Background (CMB) is located at multipole
number $l \sim$ 200. This suggests that the Universe is flat (or, at least,
nearly flat) consistent with the predictions of standard inflationary theories
that $\Omega_{tot}$ = 1 (de Bernardis et al. 2000, Hanany et al. 2000).
Moreover, the results of recent intermediate redshift supernova type Ia (SNIa)
surveys 
indicate that the cosmological constant is non-zero (Riess et al. 1998, 
Perlmutter et al. 1999). Taken together, the above 
results indicate that ($\Omega_M,\Omega_{\Lambda}) \sim (0.3,0.7)$, where
$\Omega_M$ is the present matter density parameter and $\Omega_{\Lambda}$ is
the density parameter corresponding to the cosmological constant $\Lambda$.

The interpretations given above of the observational results are, of course, a
matter of debate: if recombination is delayed due to, e.g., sources of
Ly$\alpha$ resonance radiation, such as stars or active galactic nuclei
(Peebles et al. 2000), or slow decay of dark matter (Doroshkevich \& Naselsky
2000, Naselsky et al. 2001) the observed position of the first Doppler peak
may actually indicate that the Universe is open (Peebles et al. 2000).
Furthermore, SNIa may not be standard candles and effects of dust absorption
(like a redshift dependent extinction law) may not have been properly 
accounted for (S. Toft, private communication). These complications could
affect the determinations of $\Omega_M$ and $\Omega_{\Lambda}$ from the CMB
and SNIa results.   

Clearly, completely independent ways of determining $\Omega_M$ and 
$\Omega_{\Lambda}$ are highly desirable. One such way is to observationally 
determine the mass function of galaxy clusters at present as well as at 
redshifts up to $z \sim 1-2$: 
\begin{figure}[t]
{\centering 
\leavevmode
\psfig{file=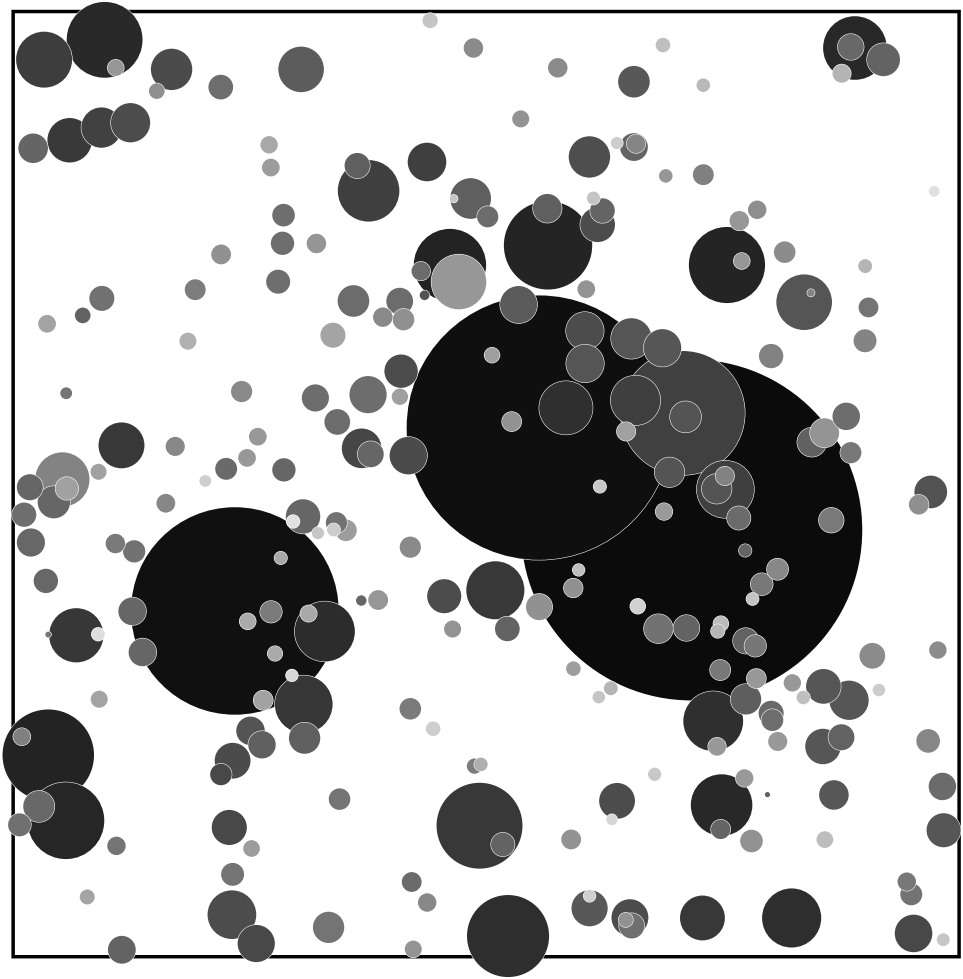,width=0.95\textwidth}}
\caption{The 247 clusters with $M \ge 10^{14}$ M$_{\odot}$ and $0<z\le1.4$ in 
a 
2.9x2.9 deg$^2$ pencil beam (almost) randomly selected from a SCDM simulation.
The grey-scale coding is such that brightness increases with redshift
- see text for details.} 
\end{figure}
\begin{figure}[t]
{\centering 
\leavevmode
\psfig{file=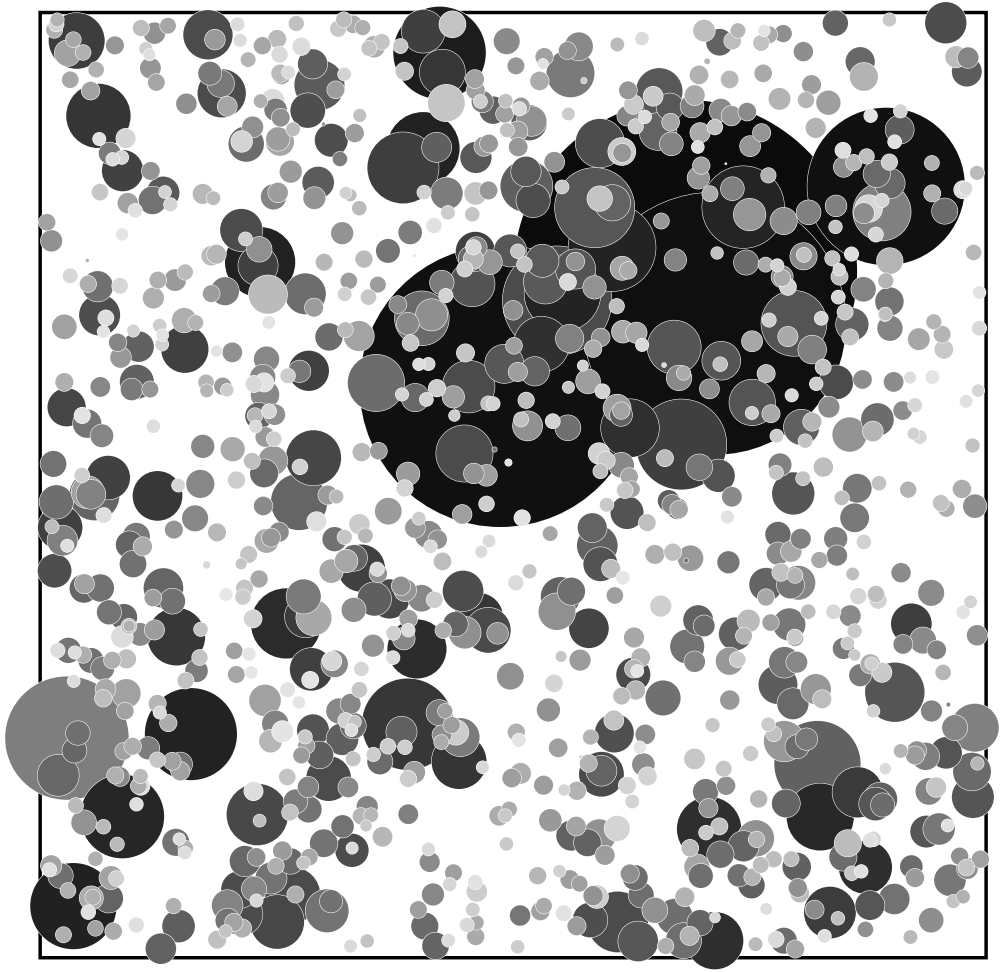,width=0.95\textwidth}}
\caption{Same pencil beam as in Figure 1, but for a $\Lambda$CDM simulation,
showing the 1016 clusters with $M \ge 10^{14}$ M$_{\odot}$ and $0<z\le1.4$.} 
\end{figure}
\begin{figure}
{\centering 
\leavevmode
\psfig{file=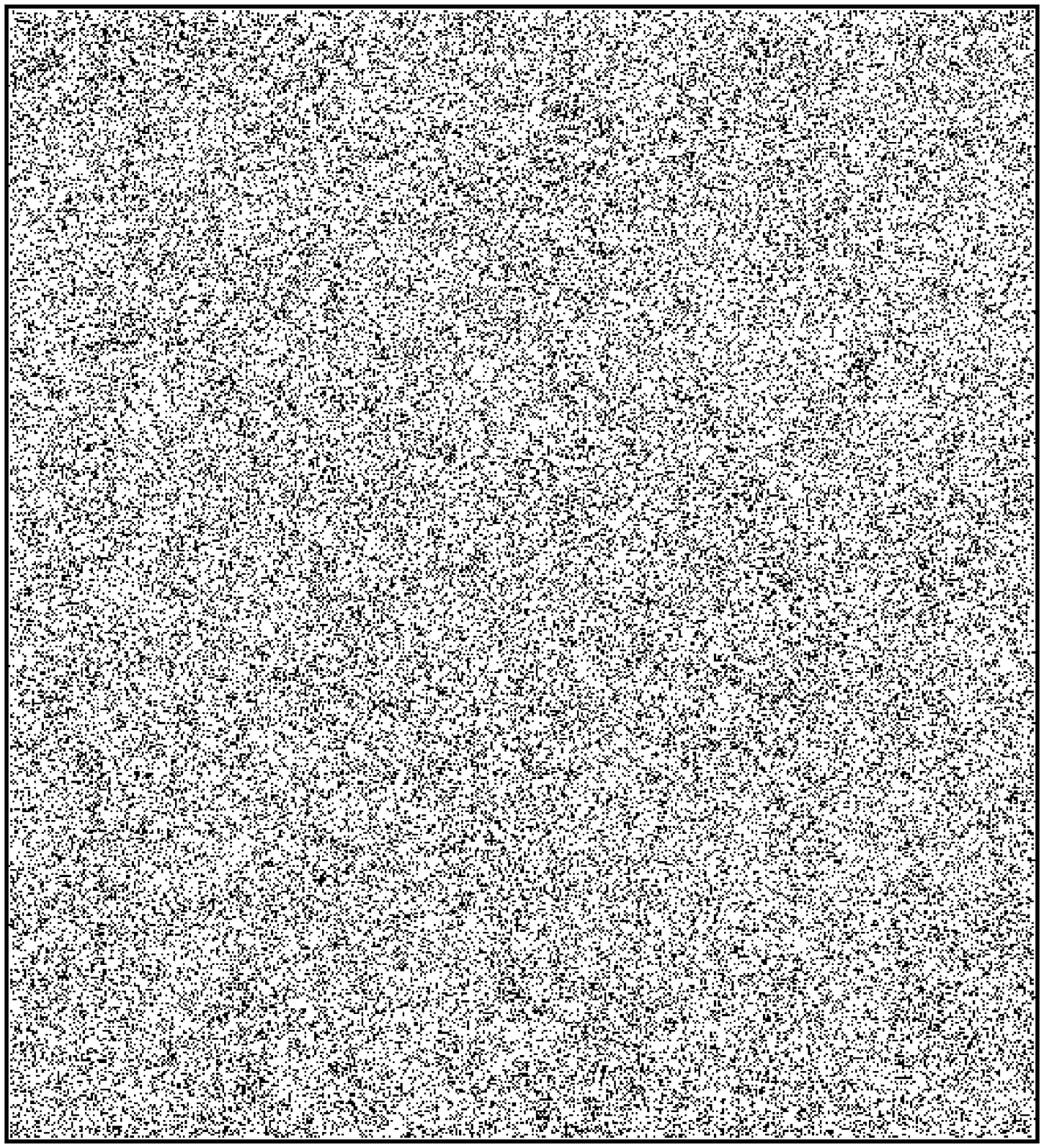,width=.49\textwidth} \hfil
\psfig{file=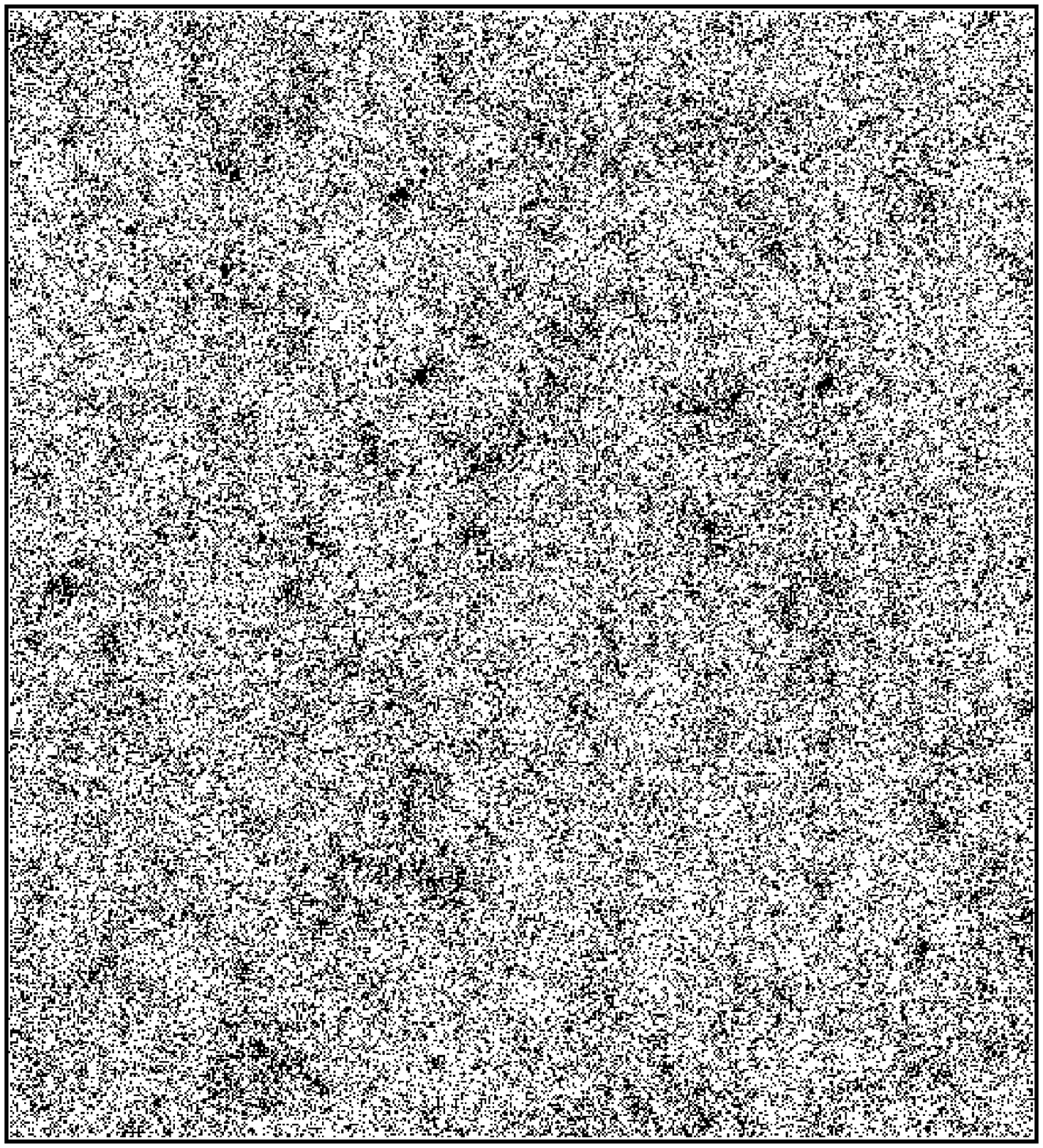,width=.49\textwidth}}
\caption{A simplified visualization of the predicted ``observed'' galaxy 
distributions corresponding to Figures 1 (left panel) and 2 (right panel), 
assuming (for simplicity) a passively evolving field galaxy luminosity 
function to represent the galaxy distribution everywhere and an apparent 
magnitude range of 12$< m <$24.}
\end{figure}

\begin{figure}[t]
{\centering 
\leavevmode
\psfig{file=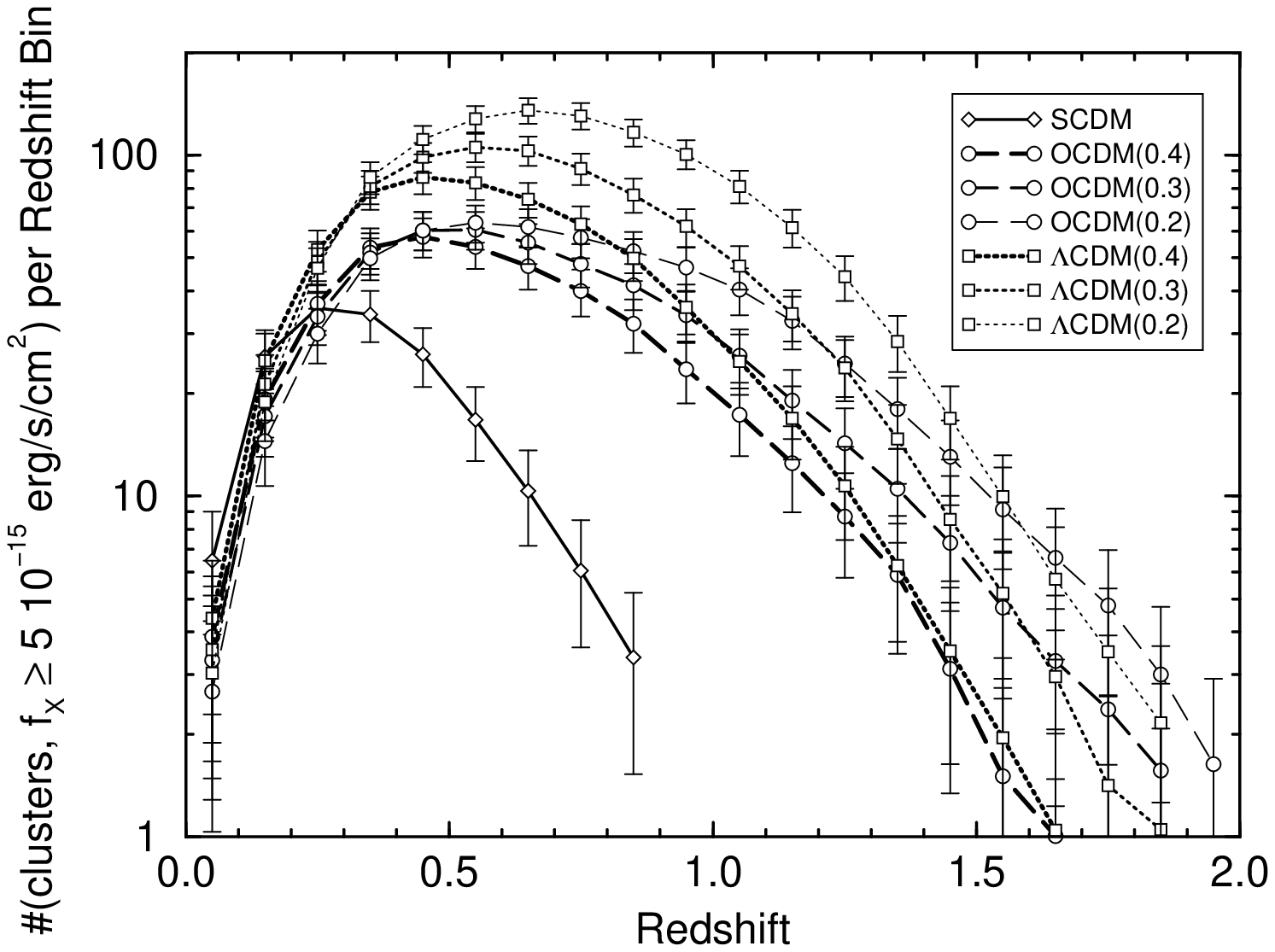,width=1.0\textwidth}}
\caption{Differential number count $dN/dz$ of X-ray clusters, brighter than the
nominal XMM-Newton flux detection limit, expected in the 64 deg$^2$ XMM-LSS 
field for various cosmologies.}
\end{figure}

\section{High redshift clusters as probes of cosmology}
The observed present-day abundance of clusters
places a strong constraint on cosmology: $\sigma_8 \Omega_M^{1/2} \simeq$ 0.5,
where $\sigma_8$ is the {\it rms} mass fluctuations on 8 $h^{-1}$ Mpc scale -
see, e.g., Eke et al. 1996 ($h$ is the present Hubble parameter in units of 
100 km/s/Mpc; the above constraint also depends weakly on $\Omega_{\Lambda}$).
This constraint is degenerate in $\Omega_M$ and $\sigma_8$, but
the degeneracy can be broken by studying the evolution of cluster abundance 
with redshift, especially for massive clusters - see, e.g., Borgani \& Guzzo
(2001) and references therein. To illustrate this in a qualitative way we show
in Figures 1 and 2 the appearance on the sky of clusters at redshifts
$0 < z \le 1.4$ in a (almost) randomly selected 2.9x2.9 deg$^2$ pencil beam 
for a SCDM and a $\Lambda$CDM cosmology (with 
($\Omega_M,\Omega_{\Lambda},h,\sigma_8$) = (1.0,0.0,0.5,0.53) and 
(0.3,0.7,0.65,1.0) respectively). The pencil beam was selected from N-body
simulations, which are effectively $\sim2\cdot10^{10}$ particle, 
``Hubble volume''
simulations. All clusters with mass $M \ge 10^{14}$ M$_{\odot}$ are shown by
filled circles with radius equal to the apparent virial radius and grey-scale 
coding
such that brightness increases with redshift. The difference between the two
figures is striking, and for the $\Lambda$CDM cosmology the cluster sky 
covering fraction appears to be $\ga$ 30\% out to a redshift of 1.4 and hence 
even larger than the otherwise impressive $\sim20$\% for the damped Ly$\alpha$
systems to redshifts of 4-5 (one might argue, though, that for detection in
X-rays or optical/NIR it would be reasonable to use $r_{500}$, rather than
the virial radius - this would reduce the cluster sky covering fraction by 
almost a factor of 4 for the $\Lambda$CDM model).

The determination of $\Omega_M$ and $\Omega_{\Lambda}$ using clusters requires
fairly large cluster surveys ideally out to redshifts of $z \sim$1.5-2 -
see below. The Copenhagen cluster group is involved in three such surveys:
1) The ESO Imaging Survey (EIS), which is an optical/NIR survey covering, when
completed, 15 deg$^2$, 2) The XMM-Newton Large Scale Survey (XMM-LSS), which 
is an X-ray survey covering, when completed, 64 deg$^2$, and finally 3) The
Planck thermal Sunyaev-Zeldovich (tSZ) all sky cluster survey (more information
about the surveys etc. is available at our web site
http://www.astro.ku.dk/xcosmos).

\subsection{Optical/NIR}
To qualitatively illustrate the potential difficulties in optical/NIR cluster
searches we made the following little experiment: At the position of each of
the $\sim10^7$ dark matter particles in the pencil beams shown in Figures 1
and 2 we randomly drew from an appropriately normalized, passively evolving
field galaxy luminosity function (LF). If the apparent magnitude of the
``galaxy'' drawn was between 12 and 24 (the approximate magnitude limits of
the EIS survey) a point was plotted in Figure 3 (SCDM: left panel, 
$\Lambda$CDM: right panel). It is
clearly difficult to spot the 247 and 1016 clusters shown in Figures 1 and 2!

In reality the observational situation is better than it seems, mainly because
the cluster LF is significantly ``brighter'' than that of the field: The 
so-called matched filter technique (Postman et al. 1996) as implemented by the
EIS team (Olsen et al. 1999) is routinely used to detect clusters up to 
$z \approx 1.3$, with tentative analysis of followup observations strongly 
supporting a cluster at $z \sim 1.2$ (L. Olsen, private communication). 

\subsection{X-rays}
Turning to X-ray detection of clusters we show in Figure 4 the expected number
of clusters per redshift bin (0.1 in $z$) in the 64 deg$^2$ XMM-LSS field 
brighter than the nominal flux detection limit of XMM. The figure is based on
N-body simulations of a range of cosmologies, the conventional 
temperature--mass ($T_X-M$) relation and a non-evolving (present-day)
luminosity--temperature ($L_X-T_X$) relation. 
It is clear from the figure how one can distinguish between
$\Lambda$ (flat, with low $\Omega_M$), open and $\Omega_M=1$ cosmologies from 
observations in the range 0.3$\la z \la$0.8 and how one can furthermore 
determine the actual value of $\Omega_M$ by adding observations in the range
1$\la z\la$2. Though the theoretical predictions obviously depends on the above
assumptions, one should note that the dependence of the $L_X-T_X$ and $T_X-M$
relations on redshift will be observationally well constrained in the future.
Clusters have been detected in X-rays up to a current record redshift of 
$z=1.79$ (Fabian et al. 2001).
  
\subsection{Thermal Sunyaev-Zeldovich effect}
When CMB photons pass through the hot gas in a galaxy cluster they inverse 
Compton scatter off hot electrons in the gas shifting the CMB spectrum to 
slightly larger energies. This causes a CMB temperature {\it decrease} in 
the low frequency (Rayleigh-Jeans) part of the CMB spectrum of $\Delta T/T_0
\simeq -2y$, where $T_0$=2.73 K and $y$ is given by
\begin{equation}
y = \frac{k_B \sigma_T}{m_e c^2} \int n_e(l) T_e(l) dl ~~,
\end{equation} 
where $\sigma_T$ the Thomson cross section, $n_e$ the electron number density,
$T_e$ the electron temperature and the integral is along the line-of-sight.

Cluster detection through the tSZ effect offers two great advantages over
X-ray detection: a) The tSZ properties of clusters are differently sensitive
to the structure of the intracluster medium than their X-ray properties
because the local contribution to the tSZ effect is proportional to the
pressure (eq. 1) whereas the local X-ray emissivity is proportional to
the square of the density times the square root of the temperature (to first
order). The tSZ effect is thus less centrally concentrated than the X-ray
emission and much less sensitive to small-scale gas clumping. Moreover, the
tSZ ``luminosity'' is proportional to the total thermal energy content of
the cluster gas (eq. 1) which should correlate fairly tightly with the
total baryon content of the cluster and hence its total mass. So the 
``$L_{SZ}-M$'' relation would be expected to be significantly tighter than
the $L_X-M$ relation. b) The tSZ ``surface brightness'' of a cluster is
independent of its distance whereas its X-ray (and optical/NIR) surface
brightness drops as $(1+z)^4$, so that very distant clusters are comparatively
easier to detect through the tSZ effect.       

The Planck survey will be all sky (except for excluded regions near the
Galactic plane) and should detect $\sim10^4$ clusters at $z\ga0.3$,
$\sim10^3$ clusters at $z\ga1$ and $\sim10^2$ clusters at $z\ga1.5$, depending
on the cosmology, of course - the numbers given are for a 
($\Omega_M,\Omega_{\Lambda}) \sim (0.3,0.7)$ Universe.

We finally note that combination of tSZ and X-ray data offers a powerful probe
of the gas structure in clusters and in some circumstances may allow a direct
determination of their individual distances and hence the Hubble parameter
(Silk \& White 1978).

\vspace{.25cm}
\section*{Acknowledgments}
\vspace{-3mm}
JSL is grateful to F.~Matteucci and the other organizers of this conference
for giving him the opportunity to participate and contribute. We have benefited
from discussions with Stefano Borgani, Lisbeth Fogh Olsen, Laura Portinari and
Sune Toft.


\end{document}